\documentclass[a4paper, amsfonts, amssymb, amsmath, reprint, showkeys, nofootinbib, twoside, notitlepage]{revtex4-2}

\def\pra{{Phys.~Rev.~A}}        
\def\prd{{Phys.~Rev.~D}}        
\def\nat{{Nature}}              

\usepackage{amsmath,amstext}
\usepackage[T1]{fontenc}
\usepackage{amssymb}
\input{epsf}
\usepackage{graphicx}
\usepackage{ae,aecompl}
\usepackage{combelow}

\usepackage{standalone}
\usepackage{hyperref}
\usepackage{amsmath}
\usepackage{amssymb}
\usepackage{mathtools}

\DeclarePairedDelimiter\floor{\lfloor}{\rfloor}
\usepackage{bm}
\usepackage{cleveref}
\usepackage{tensor}
\usepackage{braket}
\usepackage{enumitem}
\usepackage{mhchem}
\usepackage{cancel}
\usepackage{tikz}
\usepackage{pgfplots}
\usetikzlibrary{external}
\usepackage{mathrsfs}
\usepackage{amsthm}
\usepackage{physics}

\usepackage{color}

\newcommand{\spc}{\quad \quad \quad}

\def\be{\begin{equation}}
\def\ee{\end{equation}}
\def\beq{\begin{eqnarray}}
\def\eeq{\end{eqnarray}}

\theoremstyle{definition}

\theoremstyle{theorem}

\theoremstyle{corollary}

\begin{document}
\title{Life on a closed timelike curve}
\author{L.~Gavassino}
\email{lorenzo.gavassino@vanderbilt.edu}
\affiliation{Department of Mathematics, Vanderbilt University, Nashville, TN 37211, USA}


\begin{abstract}
We study the internal dynamics of a hypothetical spaceship traveling on a close timelike curve in an axially symmetric universe. We choose the curve so that the generator of evolution in proper time is the angular momentum. Using Wigner's theorem, we prove that the energy levels internal to the spaceship must undergo spontaneous discretization. The level separation turns out to be finely tuned so that, after completing a roundtrip of the curve, all systems are back to their initial state. This implies, for example, that the memories of an observer inside the spaceship are necessarily erased by the end of the journey. More in general, if there is an increase in entropy, a Poincar\'{e} cycle will eventually reverse it by the end of the loop, forcing entropy to decrease back to its initial value. We show that such decrease in entropy is in agreement with the eigenstate thermalization hypothesis. The non-existence of time-travel paradoxes follows as a rigorous corollary of our analysis.
\end{abstract}

\maketitle

\section{Introduction}

It is often assumed that, in a universe with Closed Timelike Curves (CTCs), people can ``travel to the past''. On the surface, this seems to be an obvious implication, since (on sufficiently large scales) one may view a timelike curve as the worldline of a hypothetical spaceship traveling across the spacetime. If such curve forms a loop, the spaceship returns to its starting point, in its own past. However, to confirm that this is an actual journey to the past, we must first discuss what happens to the passengers (i.e. to macroscopic systems of particles) as they complete the roundtrip. For example, consider the following question: ``\textit{Can Alice meet her younger self at the end of the journey?}'' Answering this and similar queries ultimately boils down to determining the statistical evolution of non-equilibrium thermodynamic systems on CTCs \cite{CarrolFromEternity}, which is what we aim to do here.



An insightful thermodynamic picture was proposed in \cite{Rovelli:2019ltw}. Let us briefly review the main idea. We start by noticing that, according to our current understanding of physics, entropy increase is the only physical law\footnote{Ignoring phenomena that may depend on the chosen interpretation of quantum mechanics.} 
that enables us to draw a fundamental distinction between past and future \cite{Hartle:2004wx,Nikolic:2004pk}. This gives us a \textit{local} criterion\footnote{Here, ``local'' means that the criterion applies to regions of spacetime that are small compared to the spacetime geometry, but still large enough to host a closed thermodynamic system.} to determine the direction of time. Namely, if a thermally isolated system (see \cite[\S 11]{landau5} for an operational definition) occupies an event $x$ in spacetime, the ``future'' lightcone at $x$ is the one that points in the direction of increasing entropy of such thermodynamic system. Let us apply this principle to a CTC, assuming that the spaceship can indeed be approximated as a thermally isolated system. 

\begin{figure}[b!]
\begin{center}
\includegraphics[width=0.32\textwidth]{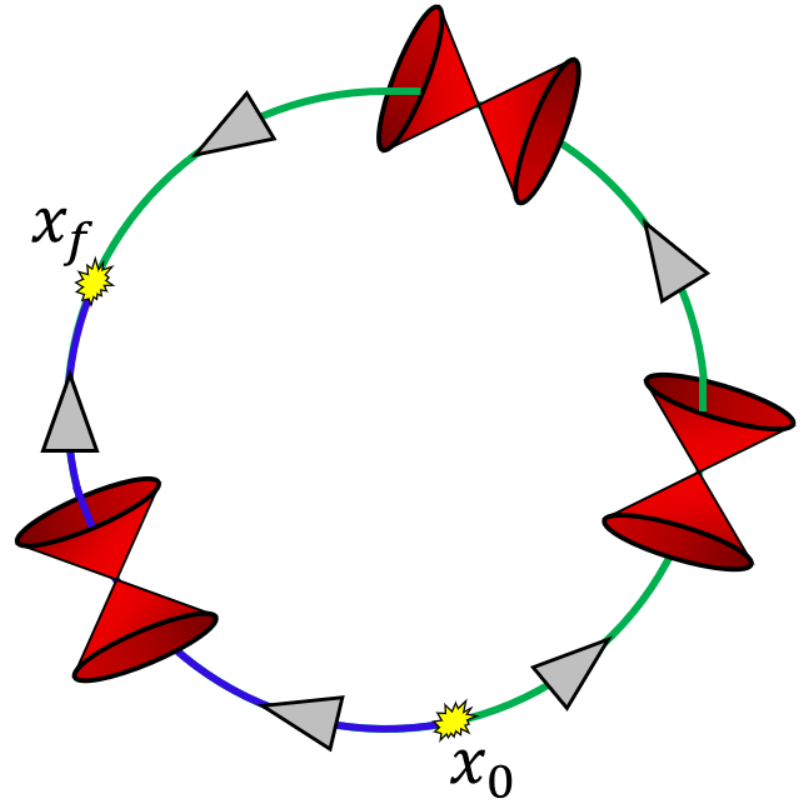}
	\caption{Entropic arrow of time (grey arrows) along a CTC. The positive sense of the arrows follows entropy increase, i.e. $dS(\text{``arrow''}){>} 0$. These time arrows must flip somewhere on the CTC, because $dS$ is an exact form, i.e. $\oint dS{=}0$. Indeed, there must be at least two points $x_0$ and $x_f$ where $dS$ changes orientation. These are the minimum and the maximum entropy events, and they mark the beginning and the end of two parallel histories (blue and green) of the same spaceship.}
	\label{fig:CTC}
	\end{center}
\end{figure}

Since a CTC is a compact set, and the entropy $S$ is a continuous function, there exists an event $x_0$ in the curve where the entropy of the spaceship is minimal, and another event $x_f$ where it is maximal\footnote{To keep the discussion simple, we assume that $x_0$ and $x_f$ are unique. The generalization to more complicated scenarios with multiple extrema is straightforward. The case where the entropy is constant along the whole CTC is not interesting for us, since this article focuses on non-equilibrium processes on CTCs.}. 
Let us then break the closed curve into two open segments, which begin in $x_0$ and end in $x_f$. Since $S(x_0)\leq S(x_f)$, the entropy grows (on average) from $x_0$ to $x_f$ on both segments. Hence, the thermodynamic time begins in $x_0$ and ends in $x_f$ on \textit{both} branches of the CTC, see Figure \ref{fig:CTC}. Based on this fact, it is argued in \cite{Rovelli:2019ltw} that, if an observer begins their journey in $x_0$, and travels counterclockwise on the green line in Figure \ref{fig:CTC}, then, after they cross the event $x_f$, we are no longer allowed to say that their subjective time ``keeps growing''. Instead, all thermodynamic processes (including biological processes such as memory formation and aging) are reversed. For this reason, strictly speaking, the story of this observer ends in $x_f$. The blue line contains an alternative story, where another (time-reversed) version of the observer travels from $x_0$ to $x_f$ clockwise. In conclusion: From the perspective of an observer traveling on the CTC, a closed timelike curve is not a loop in (entropic) time. Rather, it is two parallel entropic timelines, both of which begin and end in the same events (namely $x_0$ and $x_f$).

The purpose of the present article is to show, using quantum statistical mechanics, that the qualitative picture of \cite{Rovelli:2019ltw} outlined above is ``correct'', in the sense that it agrees with our current understanding of the laws of physics. To achieve this goal, we study the evolution of a ``spaceship'' (i.e., of a closed quantum system) traveling along a CTC that is the orbit of a local Killing vector. We will use, as our reference spacetimes, the G\"{o}del-type universes \cite{GodelPaper,Raychauduri1980}, since they have a very simple geometry. However, our main conclusions apply to a wide variety of spacetimes, notably Kerr and  \cite{Tomimatsu1973,SoenOri1996,Ori2007,Tippett:2013uwa,Fermi:2018uxf}. To be as rigorous as possible, we will ground all our claims on commonly accepted quantum mechanical principles and theorems, without invoking any assumption that may be regarded as controversial. We will verify that the thermodynamic picture of Figure \ref{fig:CTC} emerges naturally.

Let us stress that, given our purposes, a quantum approach is needed, as it provides the most practical way to prevent the insurgence of self-contradictory timelines. To see why this is important, note that, despite grandfather paradoxes are formally avoided within the understanding of Figure \ref{fig:CTC}, analogous puzzles may still arise. For example, consider an observer, called Alice, whose life begins at $x_0$, together with a coin and a notebook. Suppose that the version of Alice that evolves clockwise along the loop flips the coin and gets heads, thereby writing ``H'' on the notebook, while the other version of Alice gets tails, and writes ``T''. Which letter appears on the notebook when the two timelines meet in the event $x_f$? 

The only way out of these paradoxes is to appeal to a self-consistency principle, according to which the only histories of the Universe that actually happen are those where everything works out consistently \cite{Lewis1976,NovikovBook1983}. For example, in Alice's experiment above, it may be assumed that the coin must always give consistent outcomes on both parallel histories. In a classical world, this must be enforced as an additional constraint on the Cauchy problem \cite{Echaverria1,BACHELOT200235,Tobar:2020ybp,Friedman1990}, while in a quantum world self-consistency automatically holds as an intrinsic feature of the evolution operator \cite{Friedman1990,Politzer1992,Deutch1991,Lloyd:2010nt}. Indeed, we will show in section \ref{qmctc} that, in our spacetimes of interest, self-consistency of histories follows as a corollary of Wigner's theorem \cite{weinbergQFT_1995}, and it holds for any initial quantum state $\ket{\Psi}$. This is crucial for us, since it would be hard to make statistical mechanical claims for a system with constraints that are non-local in time.

Throughout the article, we adopt the metric signature $(-,+,+,+)$, and work in natural units: $c=\hbar=k_B=1$.

\section{Quantum Mechanics in a G\"{o}del-type universe}

In this section, we show that the energy levels of a closed system traveling along a CTC are quantized, being multiples of a fundamental discrete unit, whose value is inversely proportional to the length of the curve. Since the present article is not concerned with the backreaction of the system on the spacetime geometry, we treat the spacetime metric as a fixed classical background, and assume that the quantum system has very small mass.

\vspace{-0.2cm}
\subsection{Brief description of the metric}\label{sectioniia}
\vspace{-0.2cm}

A G\"{o}del-type universe \cite{GodelPaper,Raychauduri1980,Lobo:2008leb} is a rotating universe that admits CTCs. We work in a cylindrical coordinate system $\{t,\phi,r,z \}$, where $t\in \mathbb{R}$ is a time coordinate, $\phi=[0,2\pi]$ is the cylindrical angle (with $0 \equiv 2\pi$), $r \geq 0$ is the distance from the cylindrical axis, and $z\in \mathbb{R}$ parametrizes the axis. For the physics we are interested in, it will suffice to postulate the following general metric:
\begin{equation}\label{godeltyoe}
    g(\partial_\mu, \partial_\nu)=
\begin{bmatrix}
A(r) & B(r) & 0 & 0 \\
B(r) & C(r) & 0 & 0 \\
0 & 0 & 1 & 0 \\
0 & 0 & 0 & 1 \\
\end{bmatrix} \, ,  \quad \text{with } \, \partial_\mu =
\begin{pmatrix}
\partial_t \\
\partial_\phi \\
\partial_r \\
\partial_z \\
\end{pmatrix}\, ,
\end{equation}
where $A$, $B$, $C$ are generic functions of $r$.
G\"{o}del's metric can be expressed in the above form \cite[\S 5.7]{Hawking_Ellis_1973}, with
\begin{equation}\label{G\"{o}del}
A{=} {-}1 \, , \quad
B{=} \sqrt{2} \sinh^2 r \, , \quad 
C{=} \sinh^2 r{-}\sinh^4 r \, . 
\end{equation}
Closed timelike curves are guaranteed to exist whenever $C(r_0)<0$ for some $r_0$ \cite{Lobo:2008leb}. In fact, when this happens, $g(\partial_\phi,\partial_\phi) <0$, meaning that $\partial_\phi$ is timelike at $r_0$. Its integral curves are therefore timelike, and have the form
\begin{equation}\label{ctcwow}
    \gamma(\phi)=(\text{const},\phi,r_0,\text{const}) \, .
\end{equation}
From the identification $0 \equiv 2\pi$, we conclude that the event $\gamma(0)$ coincides with the event $\gamma(2\pi)$, meaning that the curves $\gamma(\phi)$ are indeed CTCs. In G\"{o}del's metric, $C(r)$ becomes negative when $r>\ln(1{+}\sqrt{2})$, see equation \eqref{G\"{o}del}.

\vspace{-0.2cm}
\subsection{Symmetries of the metric and their quantum realizations}\label{MJ}
\vspace{-0.2cm}

The G\"{o}del metric is known to possess 5 linearly independent Killing vectors \cite{Rebocas1983}. For the purposes of our work, only one is relevant, namely $\partial_\phi$, which is the generator of rotations around the $z$ axis in all G\"{o}del-type universes of the form \eqref{godeltyoe}. This family of transformations is an abelian one-parameter group of symmetries of the spacetime, which act as follows:
\begin{equation}\label{symmetrisa}
    \mathcal{G}_{b}(t,\phi,r,z) = (t,\phi+b,r,z) \, .
\end{equation}
If we treat the spacetime as a fixed background, the laws of physics must be invariant under all transformations $\mathcal{G}_{b}$. According to Wigner's theorem, if a quantum theory is invariant under a group of symmetries, its Hilbert space is the carrier space of a unitary representation of such group \cite[\S 2.2]{weinbergQFT_1995}. For the group \eqref{symmetrisa}, the corresponding unitary representation takes the form (by Stone's theorem \cite{Teschlbook})
\begin{equation}\label{gruppone}
U(b)= e^{iJb} \, ,
\end{equation}
for some Hermitian operator $J$. This operator is the quantum generator of rotations around the $z$-axis, and it can be identified with the angular momentum. 

A key feature of the group $\mathcal{G}_{b}$ is that, due to the identification $0 \equiv 2\pi$ for the variable $\phi$, we have the relation $\mathcal{G}_{2\pi}=1$. The group \eqref{gruppone}, being a representation of \eqref{symmetrisa}, must reflect this constraint. In particular, the action of the operator $U(2\pi)$ on a quantum state $\ket{\Psi}$ must return the same state $\ket{\Psi}$, up to a constant phase $e^{i\alpha}$ independent of $\ket{\Psi}$ \cite{weinbergQFT_1995} (if there are superselection rules, we must work within a coherent space with constant $\alpha$ \cite[\S 1-2]{Wightman2000pct}). Therefore, we have the following operatorial identity:
\begin{equation}\label{cool!}
e^{i2\pi J}=\text{const}=e^{i\alpha} \, .
\end{equation}
A more detailed proof of \eqref{cool!}, which makes explicit use of the definition of symmetry in quantum mechanics, is provided in Appendix \ref{theproof}.

\vspace{-0.2cm}
\subsection{Quantum mechanics on a CTC}\label{qmctc}
\vspace{-0.2cm}

Consider a spaceship that travels along the closed timelike curve \eqref{ctcwow}, with $r_0$ such that $C(r_0)<0$. For an observer on that spaceship, the spacetime is stationary, but the generator of time evolution is $\partial_\phi$ (and not $\partial_t$). In fact, the worldline of the spaceship is tangential to $\partial_\phi$, and a clock inside the spaceship measures proper time intervals as follows:
\begin{equation}\label{sevenone}
\Delta \tau = \int_{\phi_1}^{\phi_2} \sqrt{-g(\partial_\phi,\partial_\phi)} \, d\phi = \sqrt{-C(r_0)} \, \Delta \varphi \, .
\end{equation}
This implies that, if the spaceship is thermally isolated, it evolves in proper time $\tau$ according to the unitary operators $e^{iJ\tau/\sqrt{-C(r_0)}}$, up to an arbitrary phase. The formal proof is provided in Appendix \ref{theinham}. Thus, the Hamiltonian operator generating the spaceship's evolution is
\begin{equation}\label{chebellissimo}
    H= -\dfrac{2\pi J}{\mathcal{T}}+\text{const} \, ,
\end{equation}
where $\mathcal{T}=2\pi \sqrt{-C(r_0)}$ is the interval of proper time that the spaceship takes to perform a whole revolution around the $z$ axis. Plugging \eqref{chebellissimo} into \eqref{cool!}, and adjusting the constant in \eqref{chebellissimo} appropriately, we obtain
\begin{equation}\label{superidentity}
    e^{-iH\mathcal{T}}=1 \, .
\end{equation}
This equation implies that history ``always works out'' self-consistently. In fact, according to equation \eqref{superidentity}, after the spaceship makes a loop around the closed timelike curve, its interior must evolve back to its original state.
Note that, according to equation \eqref{superidentity}, self-consistency of history does not require that we impose constraints on the initial state. Instead, all stories are automatically self-consistent due to the operatorial identity \eqref{superidentity}, which constrains the microscopic motion of all particles.

We remark that, in order to achieve self-consistency, we did not need to invoke any exotic physics. Instead, the identity \eqref{superidentity} follows from \eqref{cool!}, which is a well-known property of the angular momentum operator. Indeed, this same quantum identity causes all objects in our world to go back to their original state (up to a global phase) after a 360$^o$ degree rotation \cite{weinbergQFT_1995}. Violating this identity would entail violating the same assumptions that underlie the foundations of the standard model of particle physics. 

\vspace{-0.2cm}
\subsection{Spontaneous discretization of the energy levels}
\vspace{-0.2cm}

Let $E$ be an eigenvalue of the Hamiltonian $H$. According to equation \eqref{superidentity}, the following fact must hold:
\begin{equation}\label{energione}
    E=\dfrac{2\pi n}{\mathcal{T}} \, , \spc n\in \mathbb{Z} \, .
\end{equation}
This means that the spectrum of $H$ is necessarily discrete, and the spacing between two arbitrary energy levels is an integer multiple of $2\pi \mathcal{T}^{-1}$. This spontaneous quantization is a necessary and sufficient condition for the self-consistency of history, since it is mathematically equivalent to \eqref{superidentity}. To understand its meaning, consider the following example. Suppose that the spaceship carries a harmonic oscillator. Self-consistency demands that, after a round-trip, the oscillator must come back to its initial position. The only way for this to happen is that its oscillation frequency $\omega$ be an integer multiple of $2\pi \mathcal{T}^{-1}$. The energy levels of the oscillator are multiples of $\omega$ (up to an irrelevant additive constant $\omega/2$), leading to \eqref{energione}.

Note that, in the limit in which the round trip takes a very long time (i.e. $\mathcal{T}\rightarrow +\infty$), the minimum quantization spacing $2\pi \mathcal{T}^{-1}$ tends to zero. For example, for a CTC whose duration is 1 year, the minimal energy spacing is $\sim 10^{-22} \, \text{eV}$. In order to have a spacing of $1 \, \text{eV}$, we need a CTC that lasts $\sim 10^{-15}$ seconds.

\vspace{-0.3cm}
\subsection{Comment on finite-size effects}\label{finitesize}
\vspace{-0.2cm}

In the above derivation, we treated the spaceship as a point-like system. However, in reality, the spaceship has a finite size, and the revolution period may be different on opposite edges of the spaceship. When this effect is relevant, all the above equations still hold, but we need to reinterpret $r_0$ as the exact location of a clock somewhere in the spaceship, and define $\mathcal{T}$ there. Then, our analysis above still applies (see Appendix \ref{appdnscale} for the proof). The only subtlety now is that we need to revise our interpretation of $\tau$. In particular, if a physical process in the spaceship lasts a time $\Delta \tau$, this does not coincide with the proper time experienced by the particles that participate in the process. Instead, it is the duration of the process as perceived by an observer located in the proximity of the clock (at $r_0$). The proof is provided in Appendix \ref{tgrsdf}. For our purposes, the present distinction is irrelevant.

\section{Spontaneous inversion of the entropic arrow of time}

In this section, we show that the thermodynamic picture illustrated in figure \ref{fig:CTC} (and introduced in \cite{Rovelli:2019ltw}) is valid (at least in axisymmetric universes), and follows directly from equation \eqref{energione}. In the following, we will model the spaceship as an idealized box with perfectly reflecting walls. This is necessary, because the second law of thermodynamics applies only to thermally isolated systems, to which we can assign a Hamiltonian \cite[\S 11]{landau5}.

\vspace{-0.2cm}
\subsection{A simple example: Spontaneous recombination of an unstable particle}
\vspace{-0.2cm}

Suppose that the content of the spaceship at $\tau =0$ amounts to one single particle, called $\Psi$, with zero kinetic energy, and which can exist in one single (normalised) quantum state $\ket{\Psi}$. Suppose that such particle is unstable, and can rapidly decay into $N\gg 1$ lighter particles, which we call $\psi$. From a statistical perspective, the probability that the daughter particles will spontaneously recombine into $\Psi$ should tend to zero in the large-$N$ limit. In fact, in the microcanonical ensemble \cite{huang_book,Callen_book}, the probability of observing the particle $\Psi$ is given by
\begin{equation}\label{probabs}
\mathcal{P}_\Psi = \dfrac{1}{1+e^{S_{N\psi}}} \sim e^{-N} \xrightarrow[]{N \rightarrow +\infty} 0 \, ,
\end{equation}
where $S_{N\psi} \sim N$ is the entropy of $N$ particles of type $\psi$. An intuitive way to see this is that it is extremely unlikely that all $N$ particles $\psi$ will spontaneously meet in the right location and with the right momenta to recreate $\Psi$. Yet, according to equation \eqref{superidentity}, after one loop around the CTC, this \textit{must} happen. Let us see how this works.

According to equation \eqref{energione}, the Hamiltonian of the system takes the general form \cite[\S 3.1, Theorem 3.7]{Teschlbook}
\begin{equation}\label{hamituzzuz}
H =\dfrac{2\pi}{\mathcal{T}} \sum_{n \in \mathbb{Z}} n P_n \, ,
\end{equation}
where $P_n=P_n^\dagger$ are orthogonal projectors, such that $P_n P_m=\delta_{nm} P_n$ and $\sum_{n \in \mathbb{Z}}  P_n=1$ \cite[Ch.1, \S 6.9]{Kato_book}. Note that the spectrum of $H$ does not need to be the whole set $2\pi \mathbb{Z}/\mathcal{T}$, because some projectors $P_n$ may vanish. The probability amplitude of observing again the particle $\Psi$ after a proper time $\tau$ has passed \cite{Fonda:1978dk} is given by
\begin{equation}\label{withone}
    \bra{\Psi}e^{-iH\tau}\ket{\Psi}= \sum_{n \in \mathbb{Z}} e^{-i 2\pi n \tau/\mathcal{T}} \bra{\Psi} P_n\ket{\Psi} \, .
\end{equation}
The quantity $\bra{\Psi} P_n\ket{\Psi}$ is the probability of obtaining the outcome $2\pi n \mathcal{T}^{-1}$ when measuring $H$ on the state $\ket{\Psi}$. We already see that equation \eqref{withone} reduces to $\bra{\Psi}e^{-iH\tau}\ket{\Psi}=1$ when $\tau{=}\mathcal{T}$. However, we are interested in understanding how the process unfolds. To this end, we assume, for clarity, that $\bra{\Psi} P_n\ket{\Psi}=1/Z$ for $n$ between $0$ and $Z{-}1$, and zero otherwise. Then, the probability of observing a particle $\Psi$ after a time $\tau$ is
\begin{figure}[!t]
\begin{center}
\includegraphics[width=0.46\textwidth]{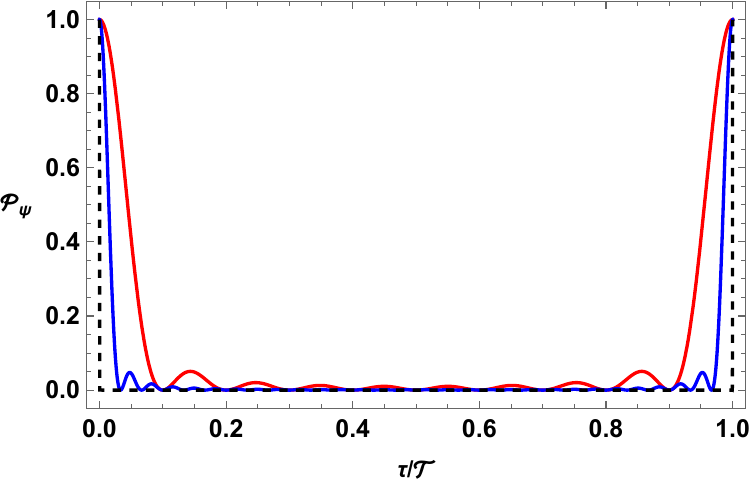}
\caption{Spontanenous decay and later recombination of an unstable particle traveling on a CTC, according to equation \eqref{ascoltami}, with $Z{=}10$ (red), $Z{=}30$ (blue), and $Z\rightarrow +\infty$ (dashed). }
	\label{fig:decay}
	\end{center}
\end{figure}
\begin{equation}\label{ascoltami}
\mathcal{P}_\Psi(\tau)= |\bra{\Psi}e^{-iH\tau}\ket{\Psi}|^2 = \dfrac{1}{Z^2} \bigg|  \dfrac{1{-}e^{-i 2\pi Z \tau/\mathcal{T}}}{1{-}e^{-i 2\pi \tau/\mathcal{T}}} \bigg|^2 \, ,   
\end{equation}
which is plotted in Figure \ref{fig:decay}. As one would expect, the particle spontaneously decays close to $\tau=0$, and it remains decayed for almost the whole journey. However, as $\tau$ approaches $\mathcal{T}$, the particle is spontaneously reconstructed, over the same time that it took for it to decay. This mechanism is a direct consequence of the discretization \eqref{energione} of the energy levels, and it does not require us to fine-tune the initial conditions. The same qualitative behavior is observed with different choices of $\bra{\Psi} P_n\ket{\Psi}$.

\vspace{-0.2cm}
\subsection{Thermodynamic evolution assuming eigenstate thermalization} 
\vspace{-0.2cm}

Let us go back to the general case, and assume that the spaceship's interior is an arbitrary closed system. Fixed an observable of interest $A$, and a normalised initial state $\ket{\Psi}$, the average value of $A$ after a time $\tau$ is
\begin{equation}
    \langle A \rangle_\tau = \bra{\Psi}e^{iH\tau}Ae^{-iH\tau}\ket{\Psi} \, .
\end{equation}
Again, it is clear that $\langle A \rangle_\mathcal{T}=\langle A \rangle_0$. Hence, if the system is initially out of thermodynamic equilibrium, it must also be such at the end of the journey (i.e. for $\tau=\mathcal{T}$). The goal is to understand what happens in between. Equation \eqref{hamituzzuz} still holds, since it follows solely from equation \eqref{energione}. Thus, we have the following decomposition:
\begin{equation}\label{gringone}
\langle A \rangle_\tau = \sum_{n,m} e^{i2\pi(n-m)\tau/\mathcal{T}} \bra{\Psi} P_n A P_m \ket{\Psi} \, .  
\end{equation}
To be able to make general qualitative statements about the evolution, let us assume for clarity that the Hamiltonian $H$ and the observable $A$ fulfill the eigenstate thermalization hypothesis \cite{DeutchETH1991,Srednicki:1994mfb,Srednicki1999,Rigol2008}. Namely, let us assume that $H$ is non-degenerate, with eigenprojectors $P_n=\ket{n}\bra{n}$, and that the following approximation holds:
\begin{equation}
\bra{n}A\ket{m} \approx \Bar{A} \delta_{nm} +  \mathcal{R}_{nm} \, ,
\end{equation}
where $\Bar{A}$ is the microcanonical average, and $\mathcal{R}_{nm}$ is a random variable with mean zero and small variance (see \cite{Srednicki1999} for more precise statements). Equation \eqref{gringone} becomes
\begin{figure}[t]
\begin{center}
\includegraphics[width=0.46\textwidth]{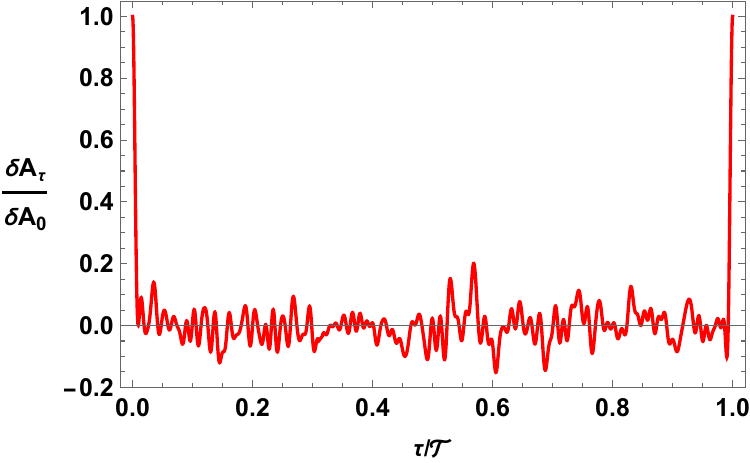}
\caption{Numerical evolution of a generic quantity $A$ in a thermodynamic system that fulfills the eigenstate thermalization hypothesis and travels along a CTC. The y-axis shows the deviation of $\langle A \rangle$ from the equilibrium value, in units of the initial deviation. The observable is assumed to be maximally out of equilibrium at $\tau=0$. The microcanonical ensemble is set to contain $100$ energy levels. Increasing the number of levels sends the relative amplitude of the thermal noise to zero.}
	\label{fig:ETH}
	\end{center}
\end{figure}
\begin{equation}\label{EhTTT}
\langle A \rangle_\tau =\Bar{A} + \sum_{n,m} e^{i2\pi(n-m)\tau/\mathcal{T}} \braket{\Psi}{n} \mathcal{R}_{nm}  \braket{m}{\Psi} \, .
\end{equation}
We can use the quantity $\delta A_\tau=\langle A\rangle_\tau{-}\Bar{A}$ as a measure of the distance of the system from thermodynamic equilibrium. In Figure \ref{fig:ETH}, we show a numerical example, for a system that starts far from equilibrium. The initial behavior is standard: The system rapidly relaxes to equilibrium, and, after that, undergoes stochastic fluctuations (which become negligible in the thermodynamic limit). However, as the end of the journey approaches, we observe a sudden large-amplitude fluctuation, which brings $\langle A \rangle_\tau$ back to where it started. The existence of large fluctuations of this kind in Hamiltonian systems was already pointed out by Poincar\'{e}, through his well-known recurrence theorem \cite{huang_book,Bocceri1957,Penrose_1979,SoaresRocha:2024afv}. In ordinary conditions, such fluctuations occur in a time that is exponentially long in the number of particles. On a CTC, this is no longer the case. The discretization of the energy levels \eqref{energione} enforces Poincar\'{e} cycles to occur over the period $\mathcal{T}$, no matter how many particles occupy the spaceship.

\vspace{-0.3cm}
\subsection{Discussion}
\vspace{-0.2cm}

Both plots \ref{fig:decay} and \ref{fig:ETH} confirm the qualitative picture outlined in \cite{Rovelli:2019ltw}, and illustrated in Figure \ref{fig:CTC}. In fact, we recall that, if the non-equilibrium deviation is not too large, the entropy of a macrostate with a given value of $A$ can be expanded as follows \cite[\S 110]{landau5}:
\begin{equation}
S(A)=S(\Bar{A})-\dfrac{1}{2} |S''(\Bar{A})| \,  (A{-}\Bar{A})^2.  
\end{equation}
Hence, we can identify the instant $\tau=0$ with the minimum-entropy event $x_0$, and the entropy grows in both time directions within a finite neighborhood of $x_0$ (see \cite{Srednicki1999} for quantitative claims). However, as we move far from $x_0$, at some point we meet an event $x_f$ where the entropy is maximal, and the second law of thermodynamics begins to turn back. In Figure \ref{fig:decay}, the event $x_f$ occurs at $\tau =\mathcal{T}/2$. In fact, around $\mathcal{T}/2$, the probability $\mathcal{P}_\Psi$ is closest to $0$, which is the thermodynamic equilibrium value, see equation \eqref{probabs}. In Figure \ref{fig:ETH}, the interior of the spaceship is in thermodynamic equilibrium across the interval $\tau/\mathcal{T} \in [0.05,0.95]$. Thus, there is no preferred direction of time in this historical age, and we can pick $x_f$ at will.

Note that the qualitative behavior in Figure \ref{fig:ETH} is essentially universal to all statistical systems fulfilling the eigenstate thermalization hypothesis and such that \eqref{superidentity} holds. In fact, all such systems must (a) evolve back to their initial state at time $\tau=\mathcal{T}$, and (b) spend most of their time in the most likely state, which is $\langle A \rangle =\Bar{A}$.

\vspace{-0.2cm}
\section{Selected topics on time travel}
\vspace{-0.2cm}

Let us discuss some general implications of our work.


\vspace{-0.2cm}
\subsection{Clocks}
\vspace{-0.2cm}

It was noted in \cite{Rovelli:2019ltw} that, on a CTC, the ticking frequency of all ``good'' clocks must be an integer multiple of $2\pi /\mathcal{T}$. In fact, self-consistency demands that, at the end of the trip, the hand of the clock must be back where it started. This leads one to the following logical implication: \textit{If the CTC's period $\mathcal{T}$ lasts a non-integer number of seconds, all second-based clocks traveling on the CTC must suffer from ``production defects'', since they will tick an integer number of times over a non-integer number of seconds}. These defects have a microscopic origin and arise due to quantum effects. Let us show this with a quick example.

Consider an Einstein's clock, namely a photon bouncing between two static mirrors with distance $D$ \cite{MerminBook}. If such mirrors are perfectly reflecting, they act as Casimir plates, discretizing the energy levels of the electromagnetic eigenmodes in the space between them. One finds that the lowest such eigenmode has excitation energy $\pi/D$ \cite{Zee2003}. However, this can be a genuine eigenmode of the evolution operator $H$ only if \eqref{energione} holds, i.e. if
\begin{equation}\label{huigunz}
    2D=\mathcal{T}/n \, , \spc n\in \mathbb{N} \, ,
\end{equation}
which is precisely the condition for the bouncing photon to be back in its initial location at the end of the loop. If \eqref{huigunz} does not hold, then $\pi/D$ cannot be an exact eigenenergy of the Hamiltonian, meaning that either the mirrors are not perfect, or they are not static (i.e. they vibrate). Either way, the Einstein clock cannot work perfectly.

\subsection{Spontaneous memory erasure}

Memory may be schematically modeled as the result of an interaction, where an object leaves traces of its initial state in the later state of a ``memory-keeper'' \cite{Hartle:2004wx}, which may be a measurement device or a living being. Let us assume, for clarity, that the state space of the spaceship can be (approximately) decomposed as the tensor product between the object's state space and the memory-keeper's state space. Then, given an orthonormal family of object's states $\ket{m}$, which the memory-keeper can distinguish through senses, memory collection can be schematically modeled as the following process\footnote{Of course, this is only a simplified picture, since the problem of how observables are measured and memory is kept is more delicate than what equations \eqref{memo1} and \eqref{memo2} may suggest. However, the main result should hold in general.}:
\begin{equation}\label{memo1}
e^{-iH \tau} \ket{m}{\otimes} \ket{\text{No knowledge}}  = \ket{\Psi_m} {\otimes} \ket{\text{``It was }m\text{''}} \, ,
\end{equation}
where $\tau$ is the duration of the interaction originating the memory. Using equation \eqref{superidentity}, we find that
\begin{equation}\label{memo2}
 e^{-iH(\mathcal{T}-\tau)} \ket{\Psi_m} {\otimes} \ket{\text{``It was }m\text{''}} =\ket{m}{\otimes} \ket{\text{No knowledge}}  \, . 
\end{equation}
Thus, \textit{any memory that is collected along the CTC will be erased by Poincar\'{e} recurrence before the end of the loop.} This confirms the claim of \cite{Rovelli:2019ltw} that the experienced time ``runs backward'' after the event $x_f$, see Figure \ref{fig:CTC}. 

\subsection{The minimum-entropy event}\label{minimumentropy}

In the introduction, we stated that, since a CTC is a compact set, there is an event $x_0$ where the entropy of the spaceship is minimal. In the proximity of such event, our macroscopic notion of causation breaks down. 
This is evident in Figures \ref{fig:decay} and \ref{fig:ETH}, where the existence of the low-entropy state at proper time $\mathcal{T}$ does not have any macroscopic cause in its near past or future. It just ``fluctuates into existence''. Indeed, any form of order that the event $x_0$ carries (including objects and people) has no logical cause that can be expressed in purely macroscopic terms. For example, if there is a book, nobody wrote it. If a person has a memory, this memory is illusory, and its content is meaningless (by human standards). This is because our notions of ``writing'' and ``forming a memory'' implicitly rely on increasing entropy \cite{CarrolFromEternity}, and there is no event with lower entropy than $x_0$.

Note that none of the above constitutes a real paradox. Complex structures (like brains) fluctuate into existence in all ergodic systems. The difference is in the waiting time $\tau_W$ before these structure forms. On CTCs, $\tau_{W} {\equiv} \mathcal{T}$. In an ergodic system with $N$ particles, $\tau_W \, {\sim} \, e^N$ \cite[\S 4.5]{huang_book}.

\subsection{Speculation: Can I meet myself from the future?}

One of the most confusing issues surrounding time travel is the possibility that an observer might interact with a version of themselves coming from the future. It is often suggested that this eventuality may give rise to retro-causality paradoxes. Let us see what quantum statistical mechanics has to say about this eventuality. We must warn the reader that the following discussion is rather speculative. Yet, we believe that it is worth pondering.


Consider a hypothetical quantum state $\ket{\text{Young Bob}}$, where a single human being lives inside the spaceship. As we evolve this state in time using $e^{-iH\tau}$, Bob's life unfolds. No matter how complicated Bob's story is, equation \eqref{superidentity} implies there is only young Bob at time $\tau{=}\mathcal{T}$:
\begin{equation}
e^{-iH \mathcal{T}}\ket{\text{Young Bob}}=\ket{\text{Young Bob}} \, .
\end{equation}
So, either old Bob dies before meeting young Bob, or he rejuvenates, becoming young Bob himself. 
Either way, we failed to arrange a meeting between two Bobs.

Let us try again. This time, we consider the quantum state $\ket{\text{Bob and Bob}}$, where two people occupy our spaceship. The first is again Bob. The second is an older version of Bob, who believes he is Bob from the future. This state of belief of ``old Bob'' is not a reliable account of his own past. In fact, based on our argument of section \ref{minimumentropy}, we know that all the memories of old Bob before some past event $x_0$ are probably false. The very existence of old Bob has no (macroscopic) causal link with the life of young Bob, because our notion of macroscopic causation breaks down around $x_0$, and order emerges with no apparent cause. Therefore, young Bob has no reason to believe that old Bob is really himself from the future. Old Bob is more likely to be an ``older clone'', who fluctuated into existence at $x_0$ for incomprehensible reasons.

\section{Conclusions}

For the first time, we have an intuitive understanding of what happens to macroscopic objects as they travel along closed timelike curves. The derivation carried out here is non-speculative, being based on the same logical procedure that underlies the unification of quantum mechanics with special relativity \cite{weinbergQFT_1995}, namely:
\begin{itemize}
\item[(i)] Fix a background spacetime of interest.
\item[(ii)] Find its symmetry group.
\item[(iii)] Construct a unitary representation of such group.
\item[(iv)] For a given observer, find which generator of the group plays the role of the Hamiltonian.
\end{itemize}
No additional postulate has been included. The resulting quantum theory was proven to be self-consistent, in the sense that all states undergo non-contradictory histories, and retro-causality paradoxes are forbidden at the quantum level (in agreement with \cite{Friedman1990,Politzer1992,Deutch1991,Lloyd:2010nt}). As suggested in \cite{Rovelli:2019ltw}, the entropic arrow of time undergoes spontaneous inversions along closed timelike curves (if the interior of the spaceship is thermally isolated). From a statistical mechanical viewpoint, this inversion is a manifestation of the Poincar\'{e} recurrence theorem, according to which, in a finite Hamiltonian system, any initial non-equilibrium deviation must fluctuate back into existence an infinite number of times (the so-called \textit{Poincar\'{e} cycles} \cite{huang_book}).

We stress that the present article does not argue in favor or against the existence of closed timelike curves. In fact, a proper discussion of the chronology protection conjecture \cite{HawkingCPC1992,ThorneBook1994xa} in G\"{o}del-type universes would require that we express the operator $J$ (defined in section \ref{MJ}) in terms of quantum fields, and that we study the backreaction of the vacuum fluctuations on the spacetime metric \cite{ThorneCTCs1992,Li1994,Li1998}, which is beyond the scope of this article\footnote{Indeed, CTCs are known to generate pathologies in interacting quantum field theories (see \cite{Friedman:2006tea} for a review). The reader can see \cite{Caldarelli:2001iq,Caldarelli:2004mz,Raeymaekers:2009ij} for derivations of the chronology protection conjecture within string theory and holography.}. Rather, the take-home message is that, in a hypothetical universe with CTCs, time travel would not take place in the form that is usually depicted in science fiction. In fact, on CTCs, thermal fluctuations destroy macroscopic causation, and erase all memories (as conjectured in \cite{Nguyen:2024efh}). As it often happens, Nature is more creative than us.

Finally, we stress again that our main results are valid in an arbitrary background spacetime (including charged Kerr black holes \cite[\S 12.3]{wald}), provided that the CTC of interest is the orbit of a periodic one-parameter family of symmetries of the metric. This happens in all axisymmetric models whose rotation Killing field $\partial_\phi$ becomes timelike somewhere.

\section*{Acknowledgements}

This work was supported by a Vanderbilt's Seeding Success Grant. We thank all referees for the useful input.

\appendix

\section{A more ``in depth'' proof of \eqref{cool!}}\label{theproof}

Let us first recall what a symmetry is, in the words of Steven Weinberg \cite[\S 2.2]{weinbergQFT_1995}:

{\it A symmetry transformation is a change in our point of view that does not change the result of possible experiments. If an observer $O$ sees a system in a state represented by a ray $\mathscr{R}$ or $\mathscr{R}_1$ or $\mathscr{R}_2$..., then an equivalent observer $O'$ who looks at the same system will observe it in a different state, represented by a ray $\mathscr{R}'$ or $\mathscr{R}'_1$ or $\mathscr{R}'_2$..., respectively, but the two observers must find the same probabilities
\begin{equation}
P(\mathscr{R}\rightarrow \mathscr{R}_n)=P(\mathscr{R}'\rightarrow \mathscr{R}'_n)\, .
\end{equation}
}

In our case, we are considering the spacetime symmetry $\phi' = \phi-b$. This corresponds to a change of perspective from an observer $O$ located at $\phi{=}0$ to an equivalent observer $O'$ located at $\phi{=}b$. In the case where $b{=}2\pi$, the second observer is located at $\phi{=}2\pi$, which is \textit{the same} as $\phi{=}0$. Hence, there is no change of perspective in this case, because the observers $O$ and $O'$ coincide. Thus, we have that $\mathscr{R}=\mathscr{R}'$ and $\mathscr{R}_n=\mathscr{R}'_n$, and the identity $e^{i2\pi J}=\text{const}=e^{i\alpha}$
follows immediately from the definition of ray and the ability to make superpositions.

\vspace{-0.2cm}
\subsection{Parallel realities are incompatible with relativity...}
\vspace{-0.2cm}

Looking at the above proof,
the careful reader may spot the following potential loophole: What if we treat the observers $O$ and $O'$ as genuinely different, by assigning to them a ``winding number'' of the phase $\phi$? In this case, there might be two ``parallel'' quantum evolutions (and two parallel realities) related to the fact that $O$ and $O'$ are formally occupying the same location $\phi{=}0$ in spacetime, but carry a different winding number, and thus observe the system to be in a different quantum state. Wouldn't this invalidate the proof?

Such a scenario does not fit within standard relativity, as it would imply that two observers at the same spacetime point can disagree on the value of a Lorentz scalar at that point. It also invalidates the way we conceptualize the spacetime manifold, since according to Einstein himself a spacetime point (or \textit{event}) is uniquely identified by the list of all ``facts'' that happen in it \cite{Einstein1905}, and two observes cannot disagree on such facts.  About this, Misner, Thorne \& Wheeler \cite[\S 1.2]{MTW_book} wrote:

{\it
Happily, nature provides its own way to localize a point in spacetime, as Einstein was the first to emphasize. Characterize the point by what happens there! Give a point in spacetime the name ``event''. Where the event lies is defined as clearly and sharply as where two straws cross each other in a barn full of hay.
}

\vspace{-0.2cm}
\subsection{...or not?}
\vspace{-0.2cm}

While the above argument rules out parallel realities \textit{at the same event} within standard physics, it also allows us to define a more conservative notion of ``parallel realities'' that fits perfectly well within canonical relativity.

Suppose that $O$ is located at $\phi=0$ and $O'$ is located at $\phi=2\pi$. Assume that $O$ measures, say, the Faraday scalar $F^{\mu \nu}F_{\mu \nu}$, and gets $5$, while $O'$ measures the same scalar $F^{\mu' \nu'}F_{\mu' \nu'}$, and gets $-3$. Then, $O$ and $O'$ are \textit{not} occupying the same event in spacetime. Instead, we should interpret the disagreement between $O$ and $O'$ as a signal that $\{ \phi=0\}\not\equiv \{\phi=2\pi\}$, and the spacetime manifold is actually a multi-sheeted Riemann-type surface, where the winding number $\floor*{\phi/(2\pi)}$ determines which sheet an observer is on. As a result, the geometry of the system is now very different from the one considered in section \ref{sectioniia}, in that:
\begin{itemize}
\item[(a)] The spacetime is no longer a G\"{o}del-type universe, since it possesses multiple $\phi-$sheets, and consequently a conical singularity at $r=0$.
\item[(b)] The curve \eqref{ctcwow} is no longer a CTC, since it never goes back to where it started. Instead, it just ``winds'' around the Riemann-type surface like a helix, with ever-increasing $\phi$.
\end{itemize}
In conclusion, the parallel-reality scenario where $O {\not\equiv} O'$ is formally allowed in relativity, but it corresponds to a geometry without CTCs. Clearly, this is not the type of spacetime considered in the present article.

\vspace{-0.2cm}
\section{The Hamiltonian of the spaceship}\label{theinham}
\vspace{-0.2cm}

Fix an observable scalar field of interest, $F(x^\mu)$, and evaluate it along the CTC \eqref{ctcwow}. The result is a one-parameter family of Hermitian operators $F(\phi)$ that an observer can measure during their journey. Working in the Heisenberg picture (which is the fundamental picture in relativity, where no preferred time is given a priori), the action of the operator $J$ on $F$ is determined by the following identity (see \cite[\S 2.4]{weinbergQFT_1995}, \cite[\S 3.2]{Coleman:2018mew}, and \cite[\S 2]{Srednicki:2007qs}):
\begin{equation}\label{fieldoni}
    e^{-iJb} F(\phi) e^{iJb} = F(\phi{+}b) \, .
\end{equation}
This is the defining property of the unitary group \eqref{gruppone}, since it makes manifest the link between the operators $U(b)$ (left-hand side) and the spacetime symmetries $\mathcal{G}_{b}$ (right-hand side)\footnote{Note that \eqref{fieldoni} can directly be used to derive \eqref{cool!}. In fact, if we set $b=2\pi$ in \eqref{fieldoni}, and use the periodicity of $\phi$ together with the single-valuedness of fields, we obtain $e^{-i2\pi J} F(\phi) e^{i2\pi J} = F(\phi)$. Thus, $e^{i2\pi J}$ must commute with all local observables. If we assume that all physical observables are built out of local ones (which is the case for quantum field theory), we conclude that $e^{i2\pi J}$ must be proportional to the identity, giving again \eqref{cool!}.}. Averaging the above equation over a generic quantum state $\ket{\Psi}$, and setting $\phi=0$, we obtain
\begin{equation}\label{heis}
\bra{\Psi} e^{-iJb} F(0) e^{iJb} \ket{\Psi} 
 = \bra{\Psi} F(b) \ket{\Psi}  
\, . 
\end{equation}
This equation allows us to define the Schr\"{o}dinger picture for observers traveling along worldlines of the form \eqref{ctcwow}. In fact, on the right-hand side of \eqref{heis}, we have a quantum average expressed in the Heisenberg picture, where the state $\ket{\Psi}$ contains the whole history of the system, and the observable $F$ is evaluated on a spacetime point of interest. On the left-hand side, the observable $F$ is fixed on the ``initial time hypersurface'', and the quantum state $e^{iJb} \ket{\Psi}$ changes depending on the ``time'' $b$ where the measurement takes place. Thus, expressing $b$ in terms of the proper time $\tau {=} \sqrt{-C(r_0)} \, b$, see equation \eqref{sevenone}, and introducing Schr\"{o}dinger's notation, i.e. $F_S{:=} F(0)$ and $\ket{\Psi(\tau)}{=}e^{iJ\tau/\sqrt{-C(r_0)}} \ket{\Psi}$, we finally obtain 
\begin{equation}
\bra{\Psi(\tau)} F_S  \ket{\Psi(\tau)} 
 = \bra{\Psi} F\big(x^\mu(\tau)\big) \ket{\Psi}  
\, .     
\end{equation}
This demonstrates that the operator $e^{iJ\tau/\sqrt{-C(r)}}$ plays indeed the role of Schr\"{o}dinger's evolution operator for an observer traveling along the worldline \eqref{ctcwow}. 

\vspace{-0.2cm}

\section{Finite-size corrections}
\vspace{-0.2cm}

In this Appendix, we expand the discussion of \ref{finitesize}.

\vspace{-0.2cm}
\subsection{Scale ambiguity of Hamiltonians in General Relativity}\label{appdnscale}
\vspace{-0.2cm}

Consider a generic spacetime manifold, with a Killing vector field $\mathcal{K}$ that is locally timelike. Then, there exists a local coordinate system $\{x^0,x^1,x^2,x^3\}$ such that $\mathcal{K}=\partial_0$, where $x^0$ is a local time coordinate. Correspondingly, we can define a Hamiltonian $H$ such that $e^{-iH\theta}$ is the unitary operator representing the translation $x^0 \rightarrow x^0+\theta$.  On the other hand, if $\mathcal{K}$ is a Killing vector, then also $\Tilde{\mathcal{K}}=a\mathcal{K}$ is a Killing vector, for any spacetime-independent constant $a{>}0$. Therefore, we can define a new coordinate system $\{ \Tilde{x}^0,\Tilde{x}^1,\Tilde{x}^2,\Tilde{x}^3\}{=}\{x^0/a,x^1,x^2,x^3 \}$ such that $\partial_{\Tilde{0}}=a\partial_0=a\mathcal{K}=\Tilde{\mathcal{K}}$. The associated Hamiltonian operator $\Tilde{H}$ will then generate the transformations $e^{-i\Tilde{H}\Tilde{\theta}}$, which represent the translations $\Tilde{x}^0 \rightarrow \Tilde{x}^0+\Tilde{\theta}$. But since $\Tilde{x}^0=x^0/a$, the unitary operator $e^{-i\Tilde{H}\Tilde{\theta}}$ must coincide with the unitary operator $e^{-iHa\Tilde{\theta}}$ (for all $\Tilde{\theta}$), so that $\Tilde{H}=aH$.
We conclude that, in general relativity, there is an intrinsic ambiguity in how to define the Hamiltonian operator, related to our ability to rescale the time coordinate\footnote{In the non-relativistic limit, this ambiguity becomes the freedom of adding an arbitrary constant to the gravitational potential. Here is the proof. Write $H=m{+}H_{\text{nr}}$ ($m$ is the rest mass, and $H_{\text{nr}}$ is the non-relativistic Hamiltonian) and $a=1{+}\epsilon$. Then, $\Tilde{H}=m +H_{\text{nr}}+\epsilon m+\epsilon H_{\text{nr}}$. In the non-relativistic limit, $\epsilon$ and $H_{\text{nr}}$ tend to zero, and the term $\epsilon H_{\text{nr}}$ may be neglected, being quadratic in infinitesimals. Therefore, we can write $\Tilde{H}\approx m +\Tilde{H}_{\text{nr}}$, with $\Tilde{H}_{\text{nr}}=H_{\text{nr}}+m\epsilon$, where the latter term can be reabsorbed in the definition of the gravitational potential.}. This ambiguity is expressed by the following group of redefinitions:
\begin{equation}\label{ambiago}
\begin{cases}
\Tilde{\mathcal{K}}= a\mathcal{K} \, ,\\
\Tilde{H}= aH \, ,\\
\Tilde{x}^0= x^0/a \, .
\end{cases}
\end{equation}

So, how do we choose the ``most natural'' value of $a$? We would like our time coordinate to coincide with the proper time of a stationary observer. Equivalently, we would like to choose the scale $a$ such that $g(a\mathcal{K},a\mathcal{K}){=}{-}1$. However, in general, it may be impossible to enforce this identity simultaneously at all locations, since
\begin{equation}
\begin{split}
\partial_\alpha[g(a\mathcal{K},a\mathcal{K})]={}& a^2\nabla_\alpha (\mathcal{K}^\mu \mathcal{K}_\mu) \\
={}& 2a^2\mathcal{K}^\mu \nabla_\alpha \mathcal{K}_\mu \\
={}& -2a^2\mathcal{K}^\mu \nabla_\mu \mathcal{K}_\alpha \, ,\\
\end{split}
\end{equation}
which does not vanish if the stationary observer is accelerating. Therefore, we need to choose a \textit{preferred} stationary worldline $(x^1_p,x^2_p,x^3_p)$, where $g(a\mathcal{K},a\mathcal{K})$ equals $-1$, and this uniquely specifies the value of $a$. When we make this choice, the time coordinate will agree with the proper time measured by the observer traveling on $(x^1_p,x^2_p,x^3_p)$, but it may not be the proper time of the neighboring observers. Similarly, the associated Hamiltonian will quantify the energy measured by detectors located at $(x^1_p,x^2_p,x^3_p)$, while in other locations it will be a redshifted (or blueshifted) energy.

In section \ref{qmctc}, we employed the ambiguity \eqref{ambiago} to adjust our definition of $H$ in \eqref{chebellissimo}. In particular, we have chosen $a$ such that $g(a\partial_\phi,a\partial_\phi)=-1$ on the CTC of interest. In an extended object, one may set $g(a\partial_\phi,a\partial_\phi)=-1$ in the object's ``center'' (in the proximity of a central clock), and there will be some redshift (or blueshift) near the boundaries. However, the spectrum of $H$ will still have the form $E_n=\text{const}\times n$, since $a$ is just a constant factor.

\subsection{Gradients in the perceived time: A thought experiment}\label{tgrsdf}

The following argument is a straightforward adaptation of the derivation of the gravitational redshift formula in stationary spacetimes, see e.g. \cite[\S 5.3]{ShapiroBook}.

We assume that the support of the spaceship is contained within the spacetime region where $\partial_\phi$ is timelike. Suppose that there are two passengers, Alice and Bob, who take two different seats in the spaceship. Alice sits on the worldline $\gamma_A(\phi)=(t_A,\phi,r_A,z_A)$, and Bob sits on the worldline $\gamma_B(\phi)=(t_B,\phi,r_B,z_B)$. Alice sends two light pulses to Bob, the first time when she is on the event $(t_A,\phi_{A1},r_A,z_A)$, and the second time when she is on the event $(t_A,\phi_{A2},r_A,z_A)$. The first lightpulse travels along a null geodesic $\gamma_1$ that begins at $(t_A,\phi_{A1},r_A,z_A)$ and terminates on Bob's worldline, at some event $(t_B,\phi_{B1},r_B,t_B)$. Now, we note that $\gamma_1$ is a null geodesic if and only if $\mathcal{G}_{b}\gamma_1$ is also a null geodesic, for any $b$ \cite[\S 25.2]{MTW_book}. Therefore, since the second light pulse is generated in the event $(t_A,\phi_{A2},r_A,z_A)$, and points in the same direction as the first, it will follow the null geodesic $\gamma_2=\mathcal{G}_{\phi_{A2}-\phi_{A1}} \gamma_1$, thereby reaching Bob in the event
\begin{equation}
\begin{split}
(t_B,\phi_{B2},r_B,z_B)={}& \mathcal{G}_{\phi_{A2}-\phi_{A1}} (t_B,\phi_{B1},r_B,t_B) \\
={}& (t_B,\phi_{B1}{+}\phi_{A2}{-}\phi_{A1},r_B,t_B) \, .
\end{split}   
\end{equation}
Hence, the coordinate delay $\Delta \phi=\phi_{B2}{-}\phi_{B1}=\phi_{A2}{-}\phi_{A1}$ between the two pulses is the same for Alice and Bob. This implies that, if Alice, e.g., has lunch in the time interval between sending the two pulses, the quantity 
\begin{equation}
\begin{split}
 \Delta \tau ={}& \sqrt{-C(r_B)} \, (\phi_{A2}{-}\phi_{A1}) \\
 ={}& \sqrt{-C(r_B)} \, (\phi_{B2}{-}\phi_{B1}) \\
 ={}& \sqrt{-C(r_B)} \, \Delta \phi   
\end{split}
\end{equation}
is the duration of Alice's lunch as perceived by Bob, see equation \eqref{sevenone}. This completes our proof.


\bibliography{Biblio}

\begin{thebibliography}{56}%
\makeatletter
\providecommand \@ifxundefined [1]{%
 \@ifx{#1\undefined}
}%
\providecommand \@ifnum [1]{%
 \ifnum #1\expandafter \@firstoftwo
 \else \expandafter \@secondoftwo
 \fi
}%
\providecommand \@ifx [1]{%
 \ifx #1\expandafter \@firstoftwo
 \else \expandafter \@secondoftwo
 \fi
}%
\providecommand \natexlab [1]{#1}%
\providecommand \enquote  [1]{``#1''}%
\providecommand \bibnamefont  [1]{#1}%
\providecommand \bibfnamefont [1]{#1}%
\providecommand \citenamefont [1]{#1}%
\providecommand \href@noop [0]{\@secondoftwo}%
\providecommand \href [0]{\begingroup \@sanitize@url \@href}%
\providecommand \@href[1]{\@@startlink{#1}\@@href}%
\providecommand \@@href[1]{\endgroup#1\@@endlink}%
\providecommand \@sanitize@url [0]{\catcode `\\12\catcode `\$12\catcode `\&12\catcode `\#12\catcode `\^12\catcode `\_12\catcode `\%12\relax}%
\providecommand \@@startlink[1]{}%
\providecommand \@@endlink[0]{}%
\providecommand \url  [0]{\begingroup\@sanitize@url \@url }%
\providecommand \@url [1]{\endgroup\@href {#1}{\urlprefix }}%
\providecommand \urlprefix  [0]{URL }%
\providecommand \Eprint [0]{\href }%
\providecommand \doibase [0]{http://dx.doi.org/}%
\providecommand \selectlanguage [0]{\@gobble}%
\providecommand \bibinfo  [0]{\@secondoftwo}%
\providecommand \bibfield  [0]{\@secondoftwo}%
\providecommand \translation [1]{[#1]}%
\providecommand \BibitemOpen [0]{}%
\providecommand \bibitemStop [0]{}%
\providecommand \bibitemNoStop [0]{.\EOS\space}%
\providecommand \EOS [0]{\spacefactor3000\relax}%
\providecommand \BibitemShut  [1]{\csname bibitem#1\endcsname}%
\let\auto@bib@innerbib\@empty
\bibitem [{\citenamefont {{Carroll, Sean}}(2010)}]{CarrolFromEternity}%
  \BibitemOpen
  \bibfield  {author} {\bibinfo {author} {\bibnamefont {{Carroll, Sean}}},\ }\href@noop {} {\emph {\bibinfo {title} {{From Eternity to Here: The Quest for the Ultimate Theory of Time}}}}\ (\bibinfo  {publisher} {{Datton}},\ \bibinfo {year} {2010})\BibitemShut {NoStop}%
\bibitem [{\citenamefont {Rovelli}(2019)}]{Rovelli:2019ltw}%
  \BibitemOpen
  \bibfield  {author} {\bibinfo {author} {\bibfnamefont {C.}~\bibnamefont {Rovelli}},\ }\href@noop {} {\  (\bibinfo {year} {2019})},\ \Eprint {http://arxiv.org/abs/1912.04702} {arXiv:1912.04702 [gr-qc]} \BibitemShut {NoStop}%
\bibitem [{\citenamefont {Hartle}(2005)}]{Hartle:2004wx}%
  \BibitemOpen
  \bibfield  {author} {\bibinfo {author} {\bibfnamefont {J.~B.}\ \bibnamefont {Hartle}},\ }\href {\doibase 10.1119/1.1783900} {\bibfield  {journal} {\bibinfo  {journal} {Am. J. Phys.}\ }\textbf {\bibinfo {volume} {73}},\ \bibinfo {pages} {101} (\bibinfo {year} {2005})},\ \Eprint {http://arxiv.org/abs/gr-qc/0403001} {arXiv:gr-qc/0403001} \BibitemShut {NoStop}%
\bibitem [{\citenamefont {Nikolic}(2006)}]{Nikolic:2004pk}%
  \BibitemOpen
  \bibfield  {author} {\bibinfo {author} {\bibfnamefont {H.}~\bibnamefont {Nikolic}},\ }\href {\doibase 10.1007/s10702-006-0516-5} {\bibfield  {journal} {\bibinfo  {journal} {Found. Phys. Lett.}\ }\textbf {\bibinfo {volume} {19}},\ \bibinfo {pages} {259} (\bibinfo {year} {2006})},\ \Eprint {http://arxiv.org/abs/gr-qc/0403121} {arXiv:gr-qc/0403121} \BibitemShut {NoStop}%
\bibitem [{\citenamefont {Landau}\ and\ \citenamefont {Lifshitz}(1980)}]{landau5}%
  \BibitemOpen
  \bibfield  {author} {\bibinfo {author} {\bibfnamefont {L.}~\bibnamefont {Landau}}\ and\ \bibinfo {author} {\bibfnamefont {E.}~\bibnamefont {Lifshitz}},\ }\href@noop {} {\emph {\bibinfo {title} {Statistical Physics}}},\ \bibinfo {number} {v. 5, Third Edition}\ (\bibinfo  {publisher} {Pergamon Press},\ \bibinfo {year} {1980})\BibitemShut {NoStop}%
\bibitem [{\citenamefont {G\"odel}(1949)}]{GodelPaper}%
  \BibitemOpen
  \bibfield  {author} {\bibinfo {author} {\bibfnamefont {K.}~\bibnamefont {G\"odel}},\ }\href {\doibase 10.1103/RevModPhys.21.447} {\bibfield  {journal} {\bibinfo  {journal} {Rev. Mod. Phys.}\ }\textbf {\bibinfo {volume} {21}},\ \bibinfo {pages} {447} (\bibinfo {year} {1949})}\BibitemShut {NoStop}%
\bibitem [{\citenamefont {Raychaudhuri}\ and\ \citenamefont {Thakurta}(1980)}]{Raychauduri1980}%
  \BibitemOpen
  \bibfield  {author} {\bibinfo {author} {\bibfnamefont {A.~K.}\ \bibnamefont {Raychaudhuri}}\ and\ \bibinfo {author} {\bibfnamefont {S.~N.~G.}\ \bibnamefont {Thakurta}},\ }\href {\doibase 10.1103/PhysRevD.22.802} {\bibfield  {journal} {\bibinfo  {journal} {Phys. Rev. D}\ }\textbf {\bibinfo {volume} {22}},\ \bibinfo {pages} {802} (\bibinfo {year} {1980})}\BibitemShut {NoStop}%
\bibitem [{\citenamefont {Tomimatsu}\ and\ \citenamefont {Sato}(1973)}]{Tomimatsu1973}%
  \BibitemOpen
  \bibfield  {author} {\bibinfo {author} {\bibfnamefont {A.}~\bibnamefont {Tomimatsu}}\ and\ \bibinfo {author} {\bibfnamefont {H.}~\bibnamefont {Sato}},\ }\href {\doibase 10.1143/PTP.50.95} {\bibfield  {journal} {\bibinfo  {journal} {Progress of Theoretical Physics}\ }\textbf {\bibinfo {volume} {50}},\ \bibinfo {pages} {95} (\bibinfo {year} {1973})},\ \Eprint {http://arxiv.org/abs/https://academic.oup.com/ptp/article-pdf/50/1/95/5362605/50-1-95.pdf} {https://academic.oup.com/ptp/article-pdf/50/1/95/5362605/50-1-95.pdf} \BibitemShut {NoStop}%
\bibitem [{\citenamefont {Soen}\ and\ \citenamefont {Ori}(1996)}]{SoenOri1996}%
  \BibitemOpen
  \bibfield  {author} {\bibinfo {author} {\bibfnamefont {Y.}~\bibnamefont {Soen}}\ and\ \bibinfo {author} {\bibfnamefont {A.}~\bibnamefont {Ori}},\ }\href {\doibase 10.1103/PhysRevD.54.4858} {\bibfield  {journal} {\bibinfo  {journal} {Phys. Rev. D}\ }\textbf {\bibinfo {volume} {54}},\ \bibinfo {pages} {4858} (\bibinfo {year} {1996})}\BibitemShut {NoStop}%
\bibitem [{\citenamefont {Ori}(2007)}]{Ori2007}%
  \BibitemOpen
  \bibfield  {author} {\bibinfo {author} {\bibfnamefont {A.}~\bibnamefont {Ori}},\ }\href {\doibase 10.1103/PhysRevD.76.044002} {\bibfield  {journal} {\bibinfo  {journal} {Phys. Rev. D}\ }\textbf {\bibinfo {volume} {76}},\ \bibinfo {pages} {044002} (\bibinfo {year} {2007})}\BibitemShut {NoStop}%
\bibitem [{\citenamefont {Tippett}\ and\ \citenamefont {Tsang}(2013)}]{Tippett:2013uwa}%
  \BibitemOpen
  \bibfield  {author} {\bibinfo {author} {\bibfnamefont {B.~K.}\ \bibnamefont {Tippett}}\ and\ \bibinfo {author} {\bibfnamefont {D.}~\bibnamefont {Tsang}},\ }\href@noop {} {\  (\bibinfo {year} {2013})},\ \Eprint {http://arxiv.org/abs/1310.7985} {arXiv:1310.7985 [gr-qc]} \BibitemShut {NoStop}%
\bibitem [{\citenamefont {Fermi}\ and\ \citenamefont {Pizzocchero}(2018)}]{Fermi:2018uxf}%
  \BibitemOpen
  \bibfield  {author} {\bibinfo {author} {\bibfnamefont {D.}~\bibnamefont {Fermi}}\ and\ \bibinfo {author} {\bibfnamefont {L.}~\bibnamefont {Pizzocchero}},\ }\href {\doibase 10.1088/1361-6382/aace6e} {\bibfield  {journal} {\bibinfo  {journal} {Class. Quant. Grav.}\ }\textbf {\bibinfo {volume} {35}},\ \bibinfo {pages} {165003} (\bibinfo {year} {2018})},\ \Eprint {http://arxiv.org/abs/1803.08214} {arXiv:1803.08214 [gr-qc]} \BibitemShut {NoStop}%
\bibitem [{\citenamefont {Lewis}(1976)}]{Lewis1976}%
  \BibitemOpen
  \bibfield  {author} {\bibinfo {author} {\bibfnamefont {D.~K.}\ \bibnamefont {Lewis}},\ }\href@noop {} {\bibfield  {journal} {\bibinfo  {journal} {American Philosophical Quarterly}\ }\textbf {\bibinfo {volume} {13}},\ \bibinfo {pages} {145} (\bibinfo {year} {1976})}\BibitemShut {NoStop}%
\bibitem [{\citenamefont {{Novikov}}(1983)}]{NovikovBook1983}%
  \BibitemOpen
  \bibfield  {author} {\bibinfo {author} {\bibfnamefont {I.~D.}\ \bibnamefont {{Novikov}}},\ }\href@noop {} {\emph {\bibinfo {title} {{Evolution of the universe}}}}\ (\bibinfo  {publisher} {Cambridge University Press (Cambridge)},\ \bibinfo {year} {1983})\BibitemShut {NoStop}%
\bibitem [{\citenamefont {Echeverria}\ \emph {et~al.}(1991)\citenamefont {Echeverria}, \citenamefont {Klinkhammer},\ and\ \citenamefont {Thorne}}]{Echaverria1}%
  \BibitemOpen
  \bibfield  {author} {\bibinfo {author} {\bibfnamefont {F.}~\bibnamefont {Echeverria}}, \bibinfo {author} {\bibfnamefont {G.}~\bibnamefont {Klinkhammer}}, \ and\ \bibinfo {author} {\bibfnamefont {K.~S.}\ \bibnamefont {Thorne}},\ }\href {\doibase 10.1103/PhysRevD.44.1077} {\bibfield  {journal} {\bibinfo  {journal} {Phys. Rev. D}\ }\textbf {\bibinfo {volume} {44}},\ \bibinfo {pages} {1077} (\bibinfo {year} {1991})}\BibitemShut {NoStop}%
\bibitem [{\citenamefont {Bachelot}(2002)}]{BACHELOT200235}%
  \BibitemOpen
  \bibfield  {author} {\bibinfo {author} {\bibfnamefont {A.}~\bibnamefont {Bachelot}},\ }\href {\doibase https://doi.org/10.1016/S0021-7824(01)01229-6} {\bibfield  {journal} {\bibinfo  {journal} {Journal de Mathématiques Pures et Appliquées}\ }\textbf {\bibinfo {volume} {81}},\ \bibinfo {pages} {35} (\bibinfo {year} {2002})}\BibitemShut {NoStop}%
\bibitem [{\citenamefont {Tobar}\ and\ \citenamefont {Costa}(2020)}]{Tobar:2020ybp}%
  \BibitemOpen
  \bibfield  {author} {\bibinfo {author} {\bibfnamefont {G.}~\bibnamefont {Tobar}}\ and\ \bibinfo {author} {\bibfnamefont {F.}~\bibnamefont {Costa}},\ }\href {\doibase 10.1088/1361-6382/aba4bc} {\bibfield  {journal} {\bibinfo  {journal} {Class. Quant. Grav.}\ }\textbf {\bibinfo {volume} {37}},\ \bibinfo {pages} {205011} (\bibinfo {year} {2020})},\ \Eprint {http://arxiv.org/abs/2001.02511} {arXiv:2001.02511 [quant-ph]} \BibitemShut {NoStop}%
\bibitem [{\citenamefont {{Friedman}}\ \emph {et~al.}(1990)\citenamefont {{Friedman}}, \citenamefont {{Morris}}, \citenamefont {{Novikov}}, \citenamefont {{Echeverria}}, \citenamefont {{Klinkhammer}}, \citenamefont {{Thorne}},\ and\ \citenamefont {{Yurtsever}}}]{Friedman1990}%
  \BibitemOpen
  \bibfield  {author} {\bibinfo {author} {\bibfnamefont {J.}~\bibnamefont {{Friedman}}}, \bibinfo {author} {\bibfnamefont {M.~S.}\ \bibnamefont {{Morris}}}, \bibinfo {author} {\bibfnamefont {I.~D.}\ \bibnamefont {{Novikov}}}, \bibinfo {author} {\bibfnamefont {F.}~\bibnamefont {{Echeverria}}}, \bibinfo {author} {\bibfnamefont {G.}~\bibnamefont {{Klinkhammer}}}, \bibinfo {author} {\bibfnamefont {K.~S.}\ \bibnamefont {{Thorne}}}, \ and\ \bibinfo {author} {\bibfnamefont {U.}~\bibnamefont {{Yurtsever}}},\ }\href {\doibase 10.1103/PhysRevD.42.1915} {\bibfield  {journal} {\bibinfo  {journal} {\prd}\ }\textbf {\bibinfo {volume} {42}},\ \bibinfo {pages} {1915} (\bibinfo {year} {1990})}\BibitemShut {NoStop}%
\bibitem [{\citenamefont {Politzer}(1992)}]{Politzer1992}%
  \BibitemOpen
  \bibfield  {author} {\bibinfo {author} {\bibfnamefont {H.~D.}\ \bibnamefont {Politzer}},\ }\href {\doibase 10.1103/PhysRevD.46.4470} {\bibfield  {journal} {\bibinfo  {journal} {Phys. Rev. D}\ }\textbf {\bibinfo {volume} {46}},\ \bibinfo {pages} {4470} (\bibinfo {year} {1992})}\BibitemShut {NoStop}%
\bibitem [{\citenamefont {Deutsch}(1991)}]{Deutch1991}%
  \BibitemOpen
  \bibfield  {author} {\bibinfo {author} {\bibfnamefont {D.}~\bibnamefont {Deutsch}},\ }\href {\doibase 10.1103/PhysRevD.44.3197} {\bibfield  {journal} {\bibinfo  {journal} {Phys. Rev. D}\ }\textbf {\bibinfo {volume} {44}},\ \bibinfo {pages} {3197} (\bibinfo {year} {1991})}\BibitemShut {NoStop}%
\bibitem [{\citenamefont {Lloyd}\ \emph {et~al.}(2011)\citenamefont {Lloyd} \emph {et~al.}}]{Lloyd:2010nt}%
  \BibitemOpen
  \bibfield  {author} {\bibinfo {author} {\bibfnamefont {S.}~\bibnamefont {Lloyd}} \emph {et~al.},\ }\href {\doibase 10.1103/PhysRevLett.106.040403} {\bibfield  {journal} {\bibinfo  {journal} {Phys. Rev. Lett.}\ }\textbf {\bibinfo {volume} {106}},\ \bibinfo {pages} {040403} (\bibinfo {year} {2011})},\ \Eprint {http://arxiv.org/abs/1005.2219} {arXiv:1005.2219 [quant-ph]} \BibitemShut {NoStop}%
\bibitem [{\citenamefont {Weinberg}(1995)}]{weinbergQFT_1995}%
  \BibitemOpen
  \bibfield  {author} {\bibinfo {author} {\bibfnamefont {S.}~\bibnamefont {Weinberg}},\ }\href {\doibase 10.1017/CBO9781139644167} {\emph {\bibinfo {title} {The Quantum Theory of Fields}}},\ Vol.~\bibinfo {volume} {1}\ (\bibinfo  {publisher} {Cambridge University Press},\ \bibinfo {year} {1995})\BibitemShut {NoStop}%
\bibitem [{\citenamefont {Lobo}(2008)}]{Lobo:2008leb}%
  \BibitemOpen
  \bibfield  {author} {\bibinfo {author} {\bibfnamefont {F.~S.~N.}\ \bibnamefont {Lobo}},\ }\href@noop {} {\  (\bibinfo {year} {2008})},\ \Eprint {http://arxiv.org/abs/1008.1127} {arXiv:1008.1127 [gr-qc]} \BibitemShut {NoStop}%
\bibitem [{\citenamefont {Hawking}\ and\ \citenamefont {Ellis}(1973)}]{Hawking_Ellis_1973}%
  \BibitemOpen
  \bibfield  {author} {\bibinfo {author} {\bibfnamefont {S.~W.}\ \bibnamefont {Hawking}}\ and\ \bibinfo {author} {\bibfnamefont {G.~F.~R.}\ \bibnamefont {Ellis}},\ }\href@noop {} {\emph {\bibinfo {title} {The Large Scale Structure of Space-Time}}},\ Cambridge Monographs on Mathematical Physics\ (\bibinfo  {publisher} {Cambridge University Press},\ \bibinfo {year} {1973})\BibitemShut {NoStop}%
\bibitem [{\citenamefont {Rebou\ifmmode~\mbox{\c{c}}\else \c{c}\fi{}as}\ and\ \citenamefont {Tiomno}(1983)}]{Rebocas1983}%
  \BibitemOpen
  \bibfield  {author} {\bibinfo {author} {\bibfnamefont {M.~J.}\ \bibnamefont {Rebou\ifmmode~\mbox{\c{c}}\else \c{c}\fi{}as}}\ and\ \bibinfo {author} {\bibfnamefont {J.}~\bibnamefont {Tiomno}},\ }\href {\doibase 10.1103/PhysRevD.28.1251} {\bibfield  {journal} {\bibinfo  {journal} {Phys. Rev. D}\ }\textbf {\bibinfo {volume} {28}},\ \bibinfo {pages} {1251} (\bibinfo {year} {1983})}\BibitemShut {NoStop}%
\bibitem [{\citenamefont {Teschl}(2009)}]{Teschlbook}%
  \BibitemOpen
  \bibfield  {author} {\bibinfo {author} {\bibfnamefont {G.}~\bibnamefont {Teschl}},\ }\href {https://isidore.co/misc/Physics%20papers%20and%20books/Mathematics/Mathematical%20Methods%20in%20Quantum%20Mechanics%20(Teschl).pdf} {\emph {\bibinfo {title} {Mathematical Methods in Quantum Mechanics With Applications to Schrodinger Operators}}}\ (\bibinfo  {publisher} {American Mathematical Society, Provence, Rhode Island},\ \bibinfo {year} {2009})\BibitemShut {NoStop}%
\bibitem [{\citenamefont {Streater}\ and\ \citenamefont {Wightman}(2000)}]{Wightman2000pct}%
  \BibitemOpen
  \bibfield  {author} {\bibinfo {author} {\bibfnamefont {R.}~\bibnamefont {Streater}}\ and\ \bibinfo {author} {\bibfnamefont {A.}~\bibnamefont {Wightman}},\ }\href {https://books.google.com/books?id=uFNwSEQ25hEC} {\emph {\bibinfo {title} {PCT, Spin and Statistics, and All that}}},\ Princeton landmarks in mathematics and physics\ (\bibinfo  {publisher} {Princeton University Press},\ \bibinfo {year} {2000})\BibitemShut {NoStop}%
\bibitem [{\citenamefont {Huang}(1987)}]{huang_book}%
  \BibitemOpen
  \bibfield  {author} {\bibinfo {author} {\bibfnamefont {K.}~\bibnamefont {Huang}},\ }\href@noop {} {\emph {\bibinfo {title} {Statistical Mechanics}}},\ \bibinfo {edition} {2nd}\ ed.\ (\bibinfo  {publisher} {John Wiley \& Sons},\ \bibinfo {year} {1987})\BibitemShut {NoStop}%
\bibitem [{\citenamefont {Callen}(1985)}]{Callen_book}%
  \BibitemOpen
  \bibfield  {author} {\bibinfo {author} {\bibfnamefont {H.~B.}\ \bibnamefont {Callen}},\ }\href {https://cds.cern.ch/record/450289} {\emph {\bibinfo {title} {{Thermodynamics and an introduction to thermostatistics; 2nd ed.}}}}\ (\bibinfo  {publisher} {Wiley},\ \bibinfo {address} {New York, NY},\ \bibinfo {year} {1985})\BibitemShut {NoStop}%
\bibitem [{\citenamefont {Kato}(1995)}]{Kato_book}%
  \BibitemOpen
  \bibfield  {author} {\bibinfo {author} {\bibfnamefont {T.}~\bibnamefont {Kato}},\ }\href@noop {} {\emph {\bibinfo {title} {{Perturbation Theory for Linear Operators}}}},\ Classics in Mathematics\ (\bibinfo  {publisher} {Springer-Verlag},\ \bibinfo {address} {Berlin},\ \bibinfo {year} {1995})\BibitemShut {NoStop}%
\bibitem [{\citenamefont {Fonda}\ \emph {et~al.}(1978)\citenamefont {Fonda}, \citenamefont {Ghirardi},\ and\ \citenamefont {Rimini}}]{Fonda:1978dk}%
  \BibitemOpen
  \bibfield  {author} {\bibinfo {author} {\bibfnamefont {L.}~\bibnamefont {Fonda}}, \bibinfo {author} {\bibfnamefont {G.~C.}\ \bibnamefont {Ghirardi}}, \ and\ \bibinfo {author} {\bibfnamefont {A.}~\bibnamefont {Rimini}},\ }\href {\doibase 10.1088/0034-4885/41/4/003} {\bibfield  {journal} {\bibinfo  {journal} {Rept. Prog. Phys.}\ }\textbf {\bibinfo {volume} {41}},\ \bibinfo {pages} {587} (\bibinfo {year} {1978})}\BibitemShut {NoStop}%
\bibitem [{\citenamefont {{Deutsch}}(1991)}]{DeutchETH1991}%
  \BibitemOpen
  \bibfield  {author} {\bibinfo {author} {\bibfnamefont {J.~M.}\ \bibnamefont {{Deutsch}}},\ }\href {\doibase 10.1103/PhysRevA.43.2046} {\bibfield  {journal} {\bibinfo  {journal} {\pra}\ }\textbf {\bibinfo {volume} {43}},\ \bibinfo {pages} {2046} (\bibinfo {year} {1991})}\BibitemShut {NoStop}%
\bibitem [{\citenamefont {Srednicki}(1994)}]{Srednicki:1994mfb}%
  \BibitemOpen
  \bibfield  {author} {\bibinfo {author} {\bibfnamefont {M.}~\bibnamefont {Srednicki}},\ }\href {\doibase 10.1103/PhysRevE.50.888} {\bibfield  {journal} {\bibinfo  {journal} {Phys. Rev. E}\ }\textbf {\bibinfo {volume} {50}} (\bibinfo {year} {1994}),\ 10.1103/PhysRevE.50.888},\ \Eprint {http://arxiv.org/abs/cond-mat/9403051} {arXiv:cond-mat/9403051} \BibitemShut {NoStop}%
\bibitem [{\citenamefont {{Srednicki}}(1999)}]{Srednicki1999}%
  \BibitemOpen
  \bibfield  {author} {\bibinfo {author} {\bibfnamefont {M.}~\bibnamefont {{Srednicki}}},\ }\href {\doibase 10.1088/0305-4470/32/7/007} {\bibfield  {journal} {\bibinfo  {journal} {Journal of Physics A Mathematical General}\ }\textbf {\bibinfo {volume} {32}},\ \bibinfo {pages} {1163} (\bibinfo {year} {1999})},\ \Eprint {http://arxiv.org/abs/cond-mat/9809360} {arXiv:cond-mat/9809360 [cond-mat.stat-mech]} \BibitemShut {NoStop}%
\bibitem [{\citenamefont {{Rigol}}\ \emph {et~al.}(2008)\citenamefont {{Rigol}}, \citenamefont {{Dunjko}},\ and\ \citenamefont {{Olshanii}}}]{Rigol2008}%
  \BibitemOpen
  \bibfield  {author} {\bibinfo {author} {\bibfnamefont {M.}~\bibnamefont {{Rigol}}}, \bibinfo {author} {\bibfnamefont {V.}~\bibnamefont {{Dunjko}}}, \ and\ \bibinfo {author} {\bibfnamefont {M.}~\bibnamefont {{Olshanii}}},\ }\href {\doibase 10.1038/nature06838} {\bibfield  {journal} {\bibinfo  {journal} {\nat}\ }\textbf {\bibinfo {volume} {452}},\ \bibinfo {pages} {854} (\bibinfo {year} {2008})},\ \Eprint {http://arxiv.org/abs/0708.1324} {arXiv:0708.1324 [cond-mat.stat-mech]} \BibitemShut {NoStop}%
\bibitem [{\citenamefont {Bocchieri}\ and\ \citenamefont {Loinger}(1957)}]{Bocceri1957}%
  \BibitemOpen
  \bibfield  {author} {\bibinfo {author} {\bibfnamefont {P.}~\bibnamefont {Bocchieri}}\ and\ \bibinfo {author} {\bibfnamefont {A.}~\bibnamefont {Loinger}},\ }\href {\doibase 10.1103/PhysRev.107.337} {\bibfield  {journal} {\bibinfo  {journal} {Phys. Rev.}\ }\textbf {\bibinfo {volume} {107}},\ \bibinfo {pages} {337} (\bibinfo {year} {1957})}\BibitemShut {NoStop}%
\bibitem [{\citenamefont {Penrose}(1979)}]{Penrose_1979}%
  \BibitemOpen
  \bibfield  {author} {\bibinfo {author} {\bibfnamefont {O.}~\bibnamefont {Penrose}},\ }\href {\doibase 10.1088/0034-4885/42/12/002} {\bibfield  {journal} {\bibinfo  {journal} {Reports on Progress in Physics}\ }\textbf {\bibinfo {volume} {42}},\ \bibinfo {pages} {1937} (\bibinfo {year} {1979})}\BibitemShut {NoStop}%
\bibitem [{\citenamefont {Soares~Rocha}\ \emph {et~al.}(2024)\citenamefont {Soares~Rocha}, \citenamefont {Gavassino},\ and\ \citenamefont {Mullins}}]{SoaresRocha:2024afv}%
  \BibitemOpen
  \bibfield  {author} {\bibinfo {author} {\bibfnamefont {G.}~\bibnamefont {Soares~Rocha}}, \bibinfo {author} {\bibfnamefont {L.}~\bibnamefont {Gavassino}}, \ and\ \bibinfo {author} {\bibfnamefont {N.}~\bibnamefont {Mullins}},\ }\href@noop {} {\  (\bibinfo {year} {2024})},\ \Eprint {http://arxiv.org/abs/2405.10878} {arXiv:2405.10878 [nucl-th]} \BibitemShut {NoStop}%
\bibitem [{\citenamefont {Mermin}(1968)}]{MerminBook}%
  \BibitemOpen
  \bibfield  {author} {\bibinfo {author} {\bibfnamefont {N.~D.}\ \bibnamefont {Mermin}},\ }\href@noop {} {\emph {\bibinfo {title} {Space and Time is Special Relativity}}}\ (\bibinfo  {publisher} {McGraw hill, New York},\ \bibinfo {year} {1968})\BibitemShut {NoStop}%
\bibitem [{\citenamefont {Zee}(2003)}]{Zee2003}%
  \BibitemOpen
  \bibfield  {author} {\bibinfo {author} {\bibfnamefont {A.}~\bibnamefont {Zee}},\ }\href@noop {} {\emph {\bibinfo {title} {{Quantum Field Theory in a Nutshell}}}},\ Nutshell handbook\ (\bibinfo  {publisher} {Princeton Univ. Press},\ \bibinfo {address} {Princeton, NJ},\ \bibinfo {year} {2003})\BibitemShut {NoStop}%
\bibitem [{\citenamefont {{Hawking}}(1992)}]{HawkingCPC1992}%
  \BibitemOpen
  \bibfield  {author} {\bibinfo {author} {\bibfnamefont {S.~W.}\ \bibnamefont {{Hawking}}},\ }\href {\doibase 10.1103/PhysRevD.46.603} {\bibfield  {journal} {\bibinfo  {journal} {\prd}\ }\textbf {\bibinfo {volume} {46}},\ \bibinfo {pages} {603} (\bibinfo {year} {1992})}\BibitemShut {NoStop}%
\bibitem [{\citenamefont {Thorne}(1994)}]{ThorneBook1994xa}%
  \BibitemOpen
  \bibfield  {author} {\bibinfo {author} {\bibfnamefont {K.~S.}\ \bibnamefont {Thorne}},\ }\href@noop {} {\emph {\bibinfo {title} {{Black holes and time warps: Einstein's outrageous legacy}}}}\ (\bibinfo  {publisher} {{W. W. Norton and Company}},\ \bibinfo {address} {New York},\ \bibinfo {year} {1994})\BibitemShut {NoStop}%
\bibitem [{\citenamefont {Thorne}(1992)}]{ThorneCTCs1992}%
  \BibitemOpen
  \bibfield  {author} {\bibinfo {author} {\bibfnamefont {K.}~\bibnamefont {Thorne}},\ }\href@noop {} {\bibfield  {journal} {\bibinfo  {journal} {``Closed Timelike Curves": Paper presented at the 13th Int. Conf. on General Relativity and Gravitation}\ } (\bibinfo {year} {1992})}\BibitemShut {NoStop}%
\bibitem [{\citenamefont {Li}(1994)}]{Li1994}%
  \BibitemOpen
  \bibfield  {author} {\bibinfo {author} {\bibfnamefont {L.-X.}\ \bibnamefont {Li}},\ }\href {\doibase 10.1103/PhysRevD.50.R6037} {\bibfield  {journal} {\bibinfo  {journal} {Phys. Rev. D}\ }\textbf {\bibinfo {volume} {50}},\ \bibinfo {pages} {R6037} (\bibinfo {year} {1994})}\BibitemShut {NoStop}%
\bibitem [{\citenamefont {Gott}\ and\ \citenamefont {Li}(1998)}]{Li1998}%
  \BibitemOpen
  \bibfield  {author} {\bibinfo {author} {\bibfnamefont {J.~R.}\ \bibnamefont {Gott}}\ and\ \bibinfo {author} {\bibfnamefont {L.-X.}\ \bibnamefont {Li}},\ }\href {\doibase 10.1103/PhysRevD.58.023501} {\bibfield  {journal} {\bibinfo  {journal} {Phys. Rev. D}\ }\textbf {\bibinfo {volume} {58}},\ \bibinfo {pages} {023501} (\bibinfo {year} {1998})}\BibitemShut {NoStop}%
\bibitem [{\citenamefont {Friedman}\ and\ \citenamefont {Higuchi}(2006)}]{Friedman:2006tea}%
  \BibitemOpen
  \bibfield  {author} {\bibinfo {author} {\bibfnamefont {J.~L.}\ \bibnamefont {Friedman}}\ and\ \bibinfo {author} {\bibfnamefont {A.}~\bibnamefont {Higuchi}},\ }\href {\doibase 10.1002/andp.200510172} {\bibfield  {journal} {\bibinfo  {journal} {Annalen Phys.}\ }\textbf {\bibinfo {volume} {15}},\ \bibinfo {pages} {109} (\bibinfo {year} {2006})},\ \Eprint {http://arxiv.org/abs/0801.0735} {arXiv:0801.0735 [gr-qc]} \BibitemShut {NoStop}%
\bibitem [{\citenamefont {Caldarelli}\ \emph {et~al.}(2001)\citenamefont {Caldarelli}, \citenamefont {Klemm},\ and\ \citenamefont {Sabra}}]{Caldarelli:2001iq}%
  \BibitemOpen
  \bibfield  {author} {\bibinfo {author} {\bibfnamefont {M.~M.}\ \bibnamefont {Caldarelli}}, \bibinfo {author} {\bibfnamefont {D.}~\bibnamefont {Klemm}}, \ and\ \bibinfo {author} {\bibfnamefont {W.~A.}\ \bibnamefont {Sabra}},\ }\href {\doibase 10.1088/1126-6708/2001/05/014} {\bibfield  {journal} {\bibinfo  {journal} {JHEP}\ }\textbf {\bibinfo {volume} {05}},\ \bibinfo {pages} {014} (\bibinfo {year} {2001})},\ \Eprint {http://arxiv.org/abs/hep-th/0103133} {arXiv:hep-th/0103133} \BibitemShut {NoStop}%
\bibitem [{\citenamefont {Caldarelli}\ \emph {et~al.}(2005)\citenamefont {Caldarelli}, \citenamefont {Klemm},\ and\ \citenamefont {Silva}}]{Caldarelli:2004mz}%
  \BibitemOpen
  \bibfield  {author} {\bibinfo {author} {\bibfnamefont {M.~M.}\ \bibnamefont {Caldarelli}}, \bibinfo {author} {\bibfnamefont {D.}~\bibnamefont {Klemm}}, \ and\ \bibinfo {author} {\bibfnamefont {P.~J.}\ \bibnamefont {Silva}},\ }\href {\doibase 10.1088/0264-9381/22/17/007} {\bibfield  {journal} {\bibinfo  {journal} {Class. Quant. Grav.}\ }\textbf {\bibinfo {volume} {22}},\ \bibinfo {pages} {3461} (\bibinfo {year} {2005})},\ \Eprint {http://arxiv.org/abs/hep-th/0411203} {arXiv:hep-th/0411203} \BibitemShut {NoStop}%
\bibitem [{\citenamefont {Raeymaekers}\ \emph {et~al.}(2010)\citenamefont {Raeymaekers}, \citenamefont {Van~den Bleeken},\ and\ \citenamefont {Vercnocke}}]{Raeymaekers:2009ij}%
  \BibitemOpen
  \bibfield  {author} {\bibinfo {author} {\bibfnamefont {J.}~\bibnamefont {Raeymaekers}}, \bibinfo {author} {\bibfnamefont {D.}~\bibnamefont {Van~den Bleeken}}, \ and\ \bibinfo {author} {\bibfnamefont {B.}~\bibnamefont {Vercnocke}},\ }\href {\doibase 10.1007/JHEP04(2010)021} {\bibfield  {journal} {\bibinfo  {journal} {JHEP}\ }\textbf {\bibinfo {volume} {04}},\ \bibinfo {pages} {021} (\bibinfo {year} {2010})},\ \Eprint {http://arxiv.org/abs/0911.3893} {arXiv:0911.3893 [hep-th]} \BibitemShut {NoStop}%
\bibitem [{\citenamefont {Nguyen}\ and\ \citenamefont {Lobo}(2024)}]{Nguyen:2024efh}%
  \BibitemOpen
  \bibfield  {author} {\bibinfo {author} {\bibfnamefont {H.~K.}\ \bibnamefont {Nguyen}}\ and\ \bibinfo {author} {\bibfnamefont {F.~S.~N.}\ \bibnamefont {Lobo}},\ }\href@noop {} {\  (\bibinfo {year} {2024})},\ \Eprint {http://arxiv.org/abs/2405.12397} {arXiv:2405.12397 [gr-qc]} \BibitemShut {NoStop}%
\bibitem [{\citenamefont {Wald}(1984)}]{wald}%
  \BibitemOpen
  \bibfield  {author} {\bibinfo {author} {\bibfnamefont {R.~M.}\ \bibnamefont {Wald}},\ }\href {https://cds.cern.ch/record/106274} {\emph {\bibinfo {title} {{General relativity}}}}\ (\bibinfo  {publisher} {Chicago Univ. Press},\ \bibinfo {address} {Chicago, IL},\ \bibinfo {year} {1984})\BibitemShut {NoStop}%
\bibitem [{\citenamefont {Einstein}(1905)}]{Einstein1905}%
  \BibitemOpen
  \bibfield  {author} {\bibinfo {author} {\bibfnamefont {A.}~\bibnamefont {Einstein}},\ }\href {\doibase 10.1002/andp.200590006} {\bibfield  {journal} {\bibinfo  {journal} {Annalen Phys.}\ }\textbf {\bibinfo {volume} {17}},\ \bibinfo {pages} {891} (\bibinfo {year} {1905})}\BibitemShut {NoStop}%
\bibitem [{\citenamefont {{Misner}}\ \emph {et~al.}(1973)\citenamefont {{Misner}}, \citenamefont {{Thorne}},\ and\ \citenamefont {{Wheeler}}}]{MTW_book}%
  \BibitemOpen
  \bibfield  {author} {\bibinfo {author} {\bibfnamefont {C.~W.}\ \bibnamefont {{Misner}}}, \bibinfo {author} {\bibfnamefont {K.~S.}\ \bibnamefont {{Thorne}}}, \ and\ \bibinfo {author} {\bibfnamefont {J.~A.}\ \bibnamefont {{Wheeler}}},\ }\href@noop {} {\emph {\bibinfo {title} {{Gravitation}}}}\ (\bibinfo  {publisher} {W.H.~Freeman and Co.},\ \bibinfo {address} {San Francisco},\ \bibinfo {year} {1973})\BibitemShut {NoStop}%
\bibitem [{\citenamefont {Coleman}(2018)}]{Coleman:2018mew}%
  \BibitemOpen
  \bibfield  {author} {\bibinfo {author} {\bibfnamefont {S.}~\bibnamefont {Coleman}},\ }\href {\doibase 10.1142/9371} {\emph {\bibinfo {title} {{Lectures of Sidney Coleman on Quantum Field Theory}}}},\ edited by\ \bibinfo {editor} {\bibfnamefont {B.~G.-g.}\ \bibnamefont {Chen}}, \bibinfo {editor} {\bibfnamefont {D.}~\bibnamefont {Derbes}}, \bibinfo {editor} {\bibfnamefont {D.}~\bibnamefont {Griffiths}}, \bibinfo {editor} {\bibfnamefont {B.}~\bibnamefont {Hill}}, \bibinfo {editor} {\bibfnamefont {R.}~\bibnamefont {Sohn}}, \ and\ \bibinfo {editor} {\bibfnamefont {Y.-S.}\ \bibnamefont {Ting}}\ (\bibinfo  {publisher} {WSP},\ \bibinfo {address} {Hackensack},\ \bibinfo {year} {2018})\BibitemShut {NoStop}%
\bibitem [{\citenamefont {Srednicki}(2007)}]{Srednicki:2007qs}%
  \BibitemOpen
  \bibfield  {author} {\bibinfo {author} {\bibfnamefont {M.}~\bibnamefont {Srednicki}},\ }\href@noop {} {\emph {\bibinfo {title} {{Quantum field theory}}}}\ (\bibinfo  {publisher} {Cambridge University Press},\ \bibinfo {year} {2007})\BibitemShut {NoStop}%
\bibitem [{\citenamefont {Shapiro}\ and\ \citenamefont {Teukolsky}(2004)}]{ShapiroBook}%
  \BibitemOpen
  \bibfield  {author} {\bibinfo {author} {\bibfnamefont {S.~L.}\ \bibnamefont {Shapiro}}\ and\ \bibinfo {author} {\bibfnamefont {S.~A.}\ \bibnamefont {Teukolsky}},\ }\href@noop {} {\emph {\bibinfo {title} {Black Holes, White Dwarfs, and Neutron Stars: The Physics of Compact Objects}}}\ (\bibinfo  {publisher} {WILEY-VCH Verlag GmbH \& Co. KGaA},\ \bibinfo {year} {2004})\BibitemShut {NoStop}%
\end{thebibliography}%

\label{lastpage}

\end{document}